\documentclass[%
 reprint,
 amsmath,amssymb,
 aps,
]{revtex4-2}
\usepackage{xcolor}
\usepackage{graphicx}
\usepackage{amsmath} 
\usepackage{dcolumn}
\usepackage{bm}

\begin{document}

\preprint{APS/123-QED}

\title{Nanoscale laser flash measurements of diffuson transport in amorphous Ge and Si}

\author{Wanyue Peng}
\author{Richard B. Wilson}%
 \email{rwilson@ucr.edu}
\affiliation{%
 Mechanical Engineering and Materials Science, University of California Riverside\\
}%





\begin{abstract}
\noindent
The thermal properties of amorphous materials have attracted significant attention due to their technological importance in electronic devices. Additionally, the disorder-induced breakdown of the phonon gas model makes vibrational transport in amorphous materials a topic of fundamental interest. In the past few decades, theoretical concepts such as propagons, diffusons, and locons have emerged to describe different types of vibrational modes in disordered solids. But experiments can struggle to accurately determine which types of vibrational states carry the majority of the heat. In the present study, we use nanoscale laser flash measurements (front/back time-domain thermoreflectance) to investigate thermal transport mechanisms in amorphous Ge and amorphous Si thin-films. We observe a nearly linear relationship between the amorphous film's thermal resistance and the film's thickness. The slope of the film's thermal resistance vs. thickness corresponds to a thickness-independent thermal conductivity of 0.4 and 0.6 W/(m-K) for a-Ge and a-Si, respectively. This result reveals that the majority of heat currents in amorphous Si and Ge thin films prepared via RF sputtering at room temperature are carried by diffusons and/or propagons with mean free paths less than a few nanometers.

\end{abstract}

\maketitle

\section{Introduction}
\noindent 
Amorphous semiconductors and insulators have attracted much interest in the past few decades due to their technological importance in optoelectronic devices \cite{singh2003advances}, potentials in thermoelectric applications, as well as unique physical properties \cite{zhou2020thermal,deangelis2019thermal,zallen2008physics,adler2013physical,cusack1988physics}.
In recent years, amorphous layers, such as Si and SiN$_x$, have been explored as interlayers between high power device heterostructures and high thermal conductivity heat spreading layers and substrates.   \cite{mu2021novel,mu2018room,mu2019high,cheng2020thermal,yates2018low}. 
The amorphous interlayer promotes bonding, which can improve thermal boundary resistance in comparison to cases with no interlayer \cite{waller2020thermal}.  However, the short mean free paths of vibrational waves in the amorphous interlayer ultimately limit the minimum possible thermal boundary resistance. Improved understanding of transport in amorphous materials, and improved techniques for studying it, could help overcome this challenge.

\noindent
\\
In comparison to their crystalline counterparts, connective orderings such as bond lengths, bond angles, and coordinations are absent in amorphous solids \cite{zallen2008physics,adler2013physical,cusack1988physics}. 
The vibrational states in disordered materials can not be described in terms of phonons due to a lack of translational symmetry \cite{hanus2021thermal,kasap2006principles}. 
To address this issue, Allen et al. developed an alternative framework to categorize vibrational states \cite{allen1999diffusons,allen1993thermal}. 
The delocalized vibrational modes with plane wave features are called \textit{propagons} \cite{allen1999diffusons,allen1993thermal,feldman1999numerical,deangelis2019thermal,hanus2021thermal}.
Extended wave packets that do not have well-defined wavevectors are called \textit{diffusons} \cite{allen1999diffusons,allen1993thermal,feldman1999numerical,deangelis2019thermal}. 
Finally, the localized vibrations are termed as \textit{locons} \cite{allen1999diffusons,allen1993thermal,feldman1999numerical,deangelis2019thermal}.
Heat currents in amorphous solids are carried by propagons and diffusons \cite{hanus2021thermal}.
Like phonons in crystalline materials, propagons in amorphous materials carry heat via wave propagation with finite mean-free paths (MFPs). Diffusons carry heat through harmonic coupling with other non-propagating diffusive modes. Each diffuson's path is a random walk of energy hops between non-propagating oscillators \cite{allen1999diffusons,allen1993thermal,feldman1999numerical,deangelis2019thermal,hanus2021thermal}. This leads to a diffuson diffusivity that depends on the frequency of successful energy hops and the jump distance between oscillators. 

\noindent
\\
An important fundamental question for the physics of solid state heat-transfer is: What vibrational modes carry heat and why? 
The amount of heat carried by propagons vs. diffusons is not believed to be a universal property that all amorphous materials have in common. Rather, it is believed that how much heat is carried by propagons vs. diffusons depends on what elements make up the compound, and how the material was synthesized \cite{braun2016size,regner2013broadband,larkin2014thermal}. Therefore, experimental methods are needed that can determine how much heat in an amorphous material is carried by propagons vs. diffusons.  

\noindent
\\
If all heat is carried by diffusons, the thermal resistance from an amorphous layer is expected to depend linearly on thickness.
However, if a significant proportion of heat is carried by long MFPs propagons, the thermal conductivity is expected to depend on thickness. In a study by Braun et al. \cite{braun2016size}, amorphous SiO$_2$ was observed to have a constant thermal conductivity of 1.1 W/(m-K) from film thicknesses from 2 nm to 2 $\mu m$, indicating diffusons and/or propagons with mean-free-paths less than 2 nm are the dominant heat carriers.
Alternatively, several studies have shown evidence that propagons can carry significant heat in amorphous Si \cite{regner2013broadband,larkin2014thermal}. 
For example, in a frequency-domain thermoreflectance (FDTR) experiment by Regner et al. \cite{regner2013broadband}, the thermal response of a 500 nm amorphous Si film was measured as a function of heating frequency.
The thermal response vs. frequency deviated from the predictions of the heat diffusion equation at MHz heating frequencies. This deviation was attributed to ballistic effects, and the authors concluded that $\sim$ 35\% of the thermal conductivity is from propagons with MFPs longer than 100 nm.

\noindent
\\
Pohl et al. \cite{pohl2002low} report that, in more than 60 materials, the ratio of an acoustic wave's mean free path to its wavelength falls into a narrow range of 10$^{-3}$ $\sim$ 10$^{-2}$ . This result holds across more than 100 studies and is independent of the measurement technique, the frequency of the acoustic wave, sample impurities, or the material's thermal history. 
However, amorphous Si, Ge, and carbon do not follow this trend. Amorphous Si, Ge, and carbon have a wavelength/mean-free-path ratio between 10$^{-4}$ and 10$^{-5}$, which is two orders of magnitude smaller than the typical value. 
This phenomenon was interpreted as a result of lattice constraints induced by the nature of four-fold coordination and may suggest propagons are more likely to be important heat carriers in these three materials.
Though Pohl et al.'s analysis indicated an equally anomalous behavior for Si, Ge, and carbon, the thermal conductivity of Si has received the most attention both experimentally and theoretically \cite{regner2013broadband,braun2016size,yang2010anomalously,zink2006thermal,liu2009high,larkin2014thermal,he2011heat,allen1999diffusons,allen1993thermal}.
In contrast, only a handful of studies have investigated amorphous Ge \cite{alvarez2008interfacial,nath1974thermal}.
\begin{figure}[!ht]
  \centering
    \includegraphics[width=0.5\textwidth]{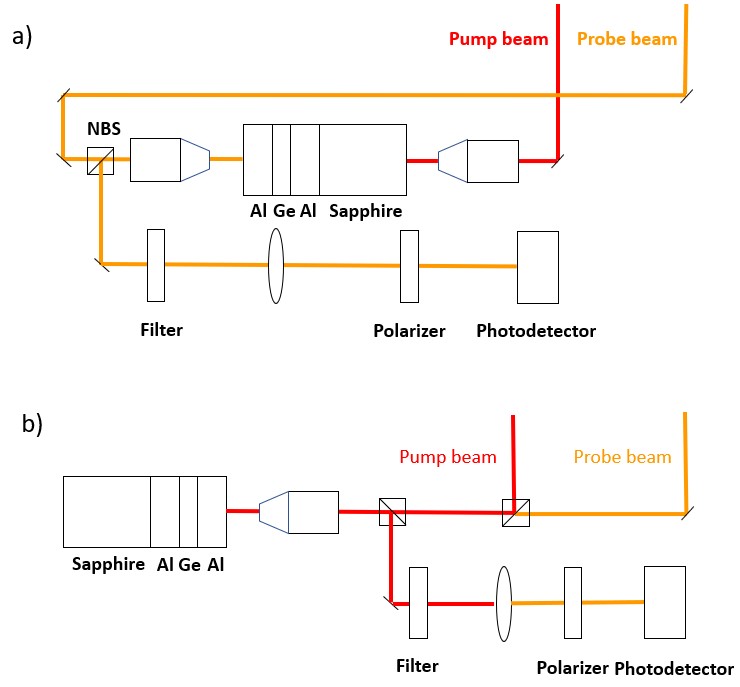}
    \caption{a) Schematic of the optical layout for our nanoscale laser flash experiments. In this front/back time domain thermoreflectance experiment, the pump beam and probe beam impinge on the opposite side of the sample. NBS is a non-polarizing beam splitter cube. The filter is a low-pass optical filter that blocks the red-shifted pump beam.  The polarizer also blocks the pump beam, which is orthogonally polarized relative to the probe beam. b) Schematic of the optical layout of the regular front/front TDTR configuration. Both the pump and probe beam impinge on the same side of the sample. }
     \label{ff_vs_fb}
\end{figure}
\begin{figure*}[!ht]
  \centering
    \includegraphics[width=0.9\textwidth]{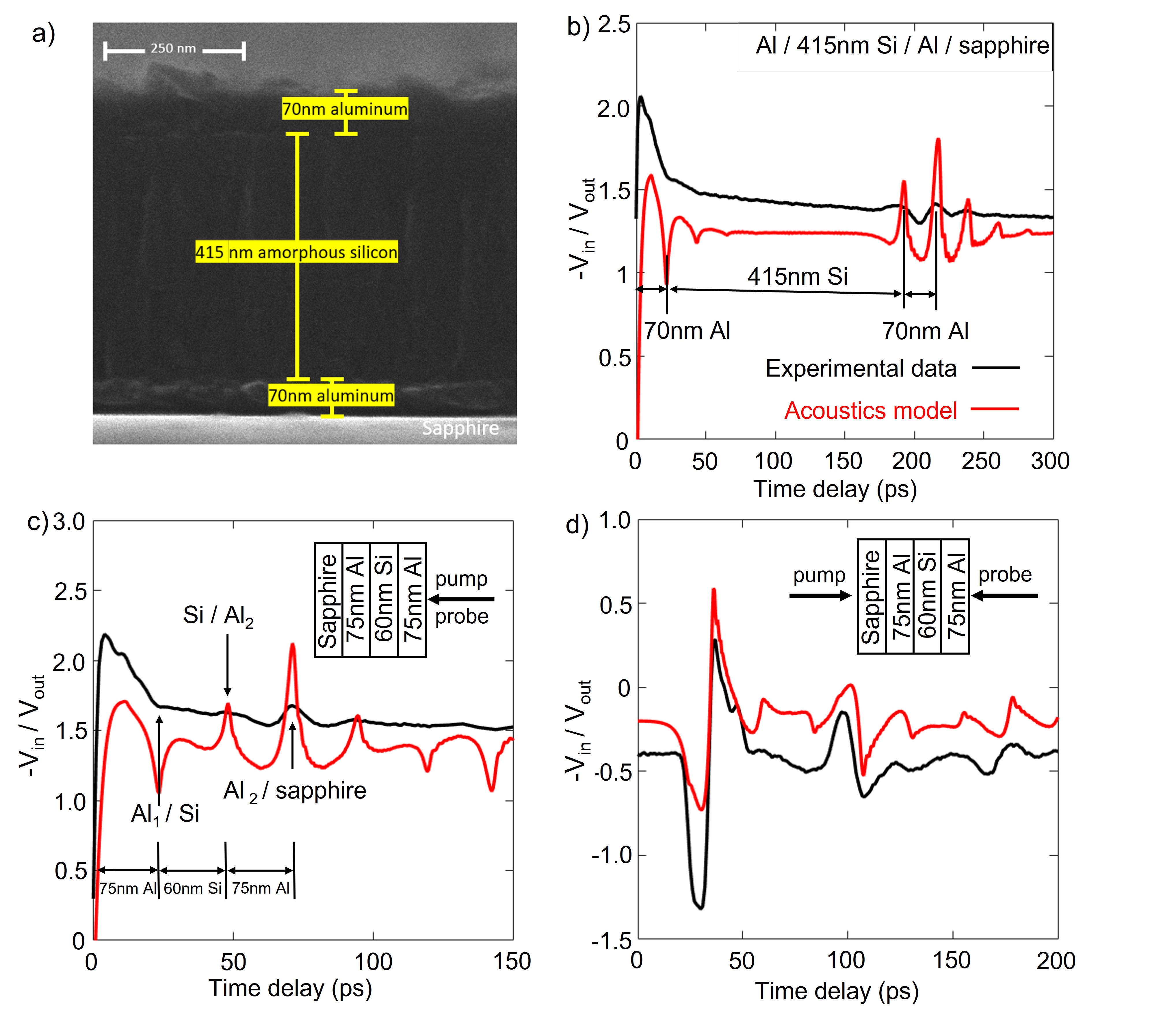}
    \caption{a) SEM thickness measurements of an Al/415nm Si/Al/sapphire stack. b) TDTR picosecond acoustic data (black) and picosecond acoustics model (red line) of 415nm Si sample shown in a). The a-Si speed of sound in the model was treated as a fit parameter. Model and data echo times agree for a speed of sound of ~3500 m/s. The echoes in the model are more pronounced than in the data because the model does not include the broadening effects of roughness. c) Comparison of front/front TDTR picosecond acoustic data and acoustic model predictions for an Al/Si/Al/sapphire sample. Echo-times predicted by the model agree with data when a-Si film thickness in the model is set $\sim$60 nm. d) Same as (c), but for a front/back configuration. The parameters deduced in c) also produce model predictions in agreement with the front/back picosecond acoustic data.}
     \label{acoustics_Si}
\end{figure*}

\noindent
\\
The aim of the present study is to shed light on the heat transport mechanisms of amorphous Ge and amorphous Si thin films with nanoscale laser flash experiments. Amorphous Ge and amorphous Si are chosen as the subject for this study because of their potentially anomalous vibrational properties in comparison to other amorphous solids \cite{pohl2002low}. The nanoscale laser flash experiments used in this study have several advantages over other experimental tools for studying thermal transport. Conventional laser flash experiments only have a time resolution on the scale of milliseconds and cannot measure thin-films \textcolor{black}{\cite{parker1961flash}}. Regular front/front time-domain thermoreflectance (TDTR) experiments \textcolor{black}{\cite{paddock1986transient,cahill2004analysis}} or frequency-domain thermoreflectance (FDTR) only measure the temperature rise at the heating surface and don't provide information on the spatial distribution of the temperature-profile. The nanoscale laser flash method used in the current study is similar to the picosecond laser flash technique used to study thin-films in Refs. \cite{taketoshi1999observation,taketoshi2001development,baba2009analysis,yagi2005analysis,oka2010thermophysical,baba2011development}. A difference in our approach and these prior picosecond laser flash studies is that we analyze the phase of the thermoreflectance signal, not just the amplitude.  The phase of the signal provides information about transport at time-scales longer than the pump-probe delay time.\\  
\noindent
\\
We find that the thermal resistance, $R_{th}$, of both amorphous Ge and Si scales nearly linearly with film thicknesses, $h$. 
The thermal resistance of the Ge film as a function of thickness is $R_{th} =$ 2.36 m-K/W$\times h+ 1.08 $ m$^2$-K/W. The thermal resistance of amorphous Si film is $R_{th} = 1.69$ m-K/W$\times h + 1.0$ m$^2$-K/W. The linear dependence of the film's thermal resistance on thickness suggests heat carriers with MFPs longer than 3 nm do not carry significant heat in amorphous Ge and Si prepared via RF sputtering at room temperature. Therefore, we conclude that diffusons are the primary heat-carrier in RF sputtered a-Ge and a-Si.
\section{Experimental methods} \label{Experimental methods}
\subsection{Materials synthesis}
\noindent
Samples of Al/Ge/Al/sapphire and Al/Si/Al/sapphire with varied thickness were prepared via sputter deposition. The Al films were DC magnetron sputter deposited. The Si and Ge layers were radio-frequency sputter deposited. The films were sputtered in an argon atmosphere on z-cut sapphire substrates. 

\noindent
\\
The Al layers serve the functions of a heater or (and) thermometer, while the Ge or Si layer is the subject of this study.
The Al, Ge, and Si sputtering targets were 5N purity. The chamber base pressure before deposition was less than 5 $\times$ 10$^{-7}$ torr. 
The Al films were sputtered at an argon pressure 3.5 $\times$ 10$^{-3}$ torr, while the Ge and Si layers were deposited at 1.5 $\times$ 10$^{-2}$ torr.
All three targets were sputtered at a power of 200 W.  
The sputtering time of each layer was set based on the deposition rate of the corresponding target.
\noindent
\begin{figure*}[!ht]
  \centering
    \includegraphics[width=0.9\textwidth]{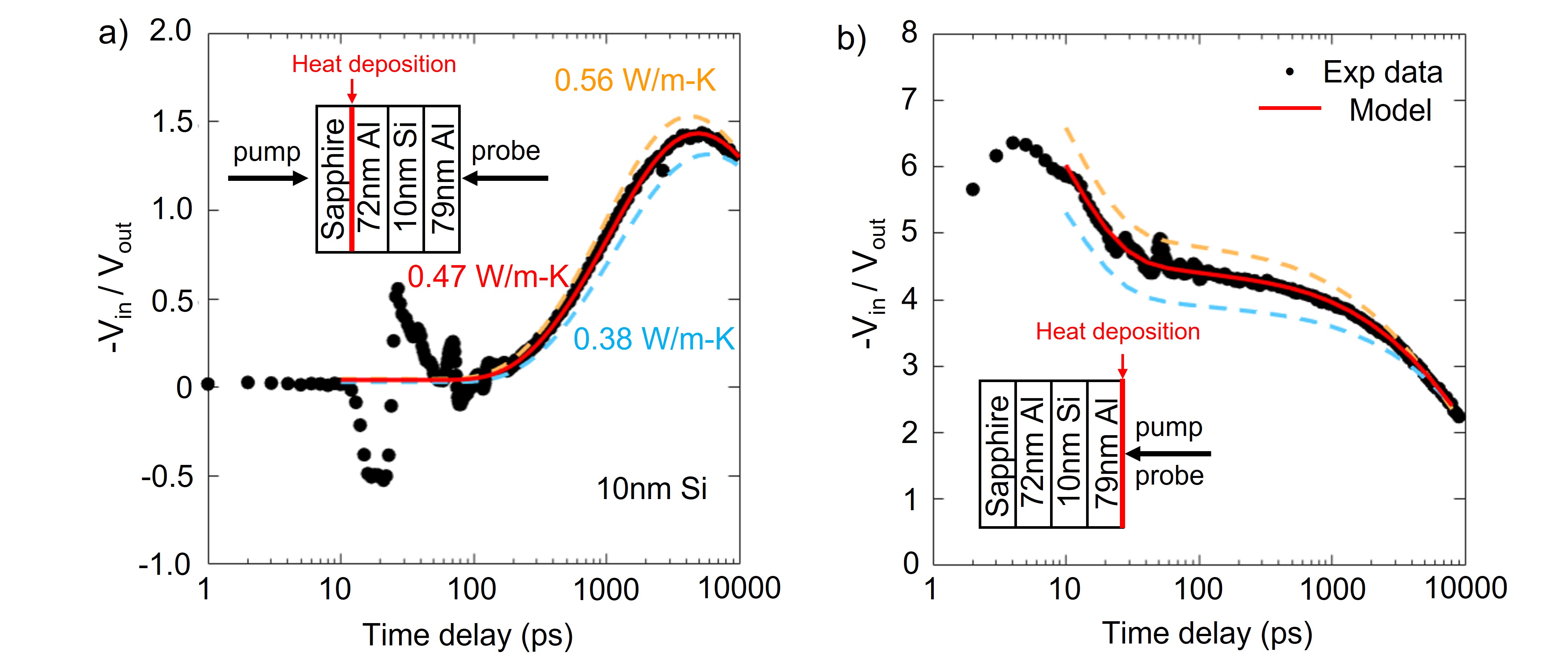}
    \caption{TDTR data and thermal model fits of Al/10nm Si/Al/sapphire in the a) front/back setup and b) front/front setup. Both measurements yield a thermal conductivity for the 10nm a-Si layer of 0.6 W/(m-K). The orange and blue dashed lines represent the fit when changing the thermal conductivity to $\pm $20\% to the current fit. }
     \label{ff_vs_fb_ratio}
\end{figure*}

\noindent
\\
We prepared four a-Ge films with thicknesses between 4 and 90 nm. We prepared five a-Si films with thicknesses between 3 and 400 nm. In addition to these nine samples, we also prepared an a-Si wedge layer with thicknesses varied between 6 and 60 nm (Al/wedge-Si/Al/sapphire). To prepare the wedge sample, a shutter was placed to cover half of the sample during the sputtering. The thickness gradient is a result of the shadowing effect from the shutter. 

\noindent
\\
The exact thicknesses of each layer were determined from picosecond acoustic signals \textcolor{black}{\cite{hohensee2012interpreting,ma2015comprehensive}} in the TDTR data, see FIG. \ref{acoustics_Si}.
The time-delay of the acoustic echoes corresponds to the time-of-flight for longitudinal acoustics waves, e.g., $2h/v$, where $h$ is the film thickness, and $v$ is the speed of sound. 
The thickness of each layer of all samples is summarized in Table 1 and 2 in the Supplementary Materials. 
To determine the speed of sound in our RF-sputtered amorphous Ge and Si layers, we collected cross-sectional scanning electron microscopy images of thick Ge and Si layers. Then, when interpreting picosecond acoustic signals on these samples, the speed of sound of the a-Si or a-Ge layer was treated as a fit parameter instead of the thickness, see FIG. \ref{ff_vs_fb_ratio}a and \ref{ff_vs_fb_ratio}b. \\
\noindent
\\
By fitting our acoustic model to the data, we deduce $v=$ 3530 m/s and 4862 m/s for a-Ge and a-Si, respectively.
These speeds of sound are comparable to literature values for amorphous films grown via sputtering.
Testardi et al. \cite{testardi1977sound} studied amorphous Ge and Si prepared by RF-sputtering and observed a speed of sound of 4400 m/s and 2850 m/s, respectively. 
Vacher et al. \cite{vacher1980attenuation} studied amorphous Si prepared by RF sputtered amorphous Si on a [110] crystalline substrate and observed a speed of sound of 3420 m/s. \\
\noindent
\\
Amorphous materials made by other approaches yield higher speeds of sound. Queen et al. \cite{queen2013excess} and Cox et al. \cite{cox1985sound}. studied amorphous Si prepared by evaporation and observed a speed of sound of 4110 m/s and 6310 m/s, respectively. Cox et al. \cite{cox1985sound} studied amorphous Ge prepared by evaporation and observed a speed of sound of 4000 m/s. Vacher et al. \cite{vacher1980attenuation} studied amorphous Ge prepared by ion bombardment and observed a speed of sound of 4290 m/s. Tan et al. \cite{tan1972elastic} studied amorphous Si prepared by ion bombardment and observed a speed of sound of 7480 m/s. Chopra et al. \cite{chopra1969thin} studied amorphous Si prepared by glow discharge and observed a speed of sound of 8430 m/s.\\
\noindent
\\
After determining the speeds of sound, we used the picosecond acoustic data for samples that were not imaged with SEM to determine the amorphous layer thickness. In FIG. \ref{acoustics_Si}c and \ref{acoustics_Si}d, we show an example of our picosecond acoustic fits for front/front and front/back measurement of a 10 nm amorphous Si film. 
%
\subsection{Thermal Characterization}
\noindent
Time-domain thermoreflectance (TDTR) measurements were performed using a pump-probe system with a Mai Tai Ti:sapphire laser with a repetition rate of 80 MHz.
In the front/front setup, both the pump and probe beam impinges on the top Al layer.
In the front/back setup, the pump beam impinges through the sapphire substrate and is absorbed by the bottom Al layer. The probe beam hits the top Al layer on the opposite side of the stack.
\begin{figure*}[!ht]
  \centering
    \includegraphics[width=0.9\textwidth]{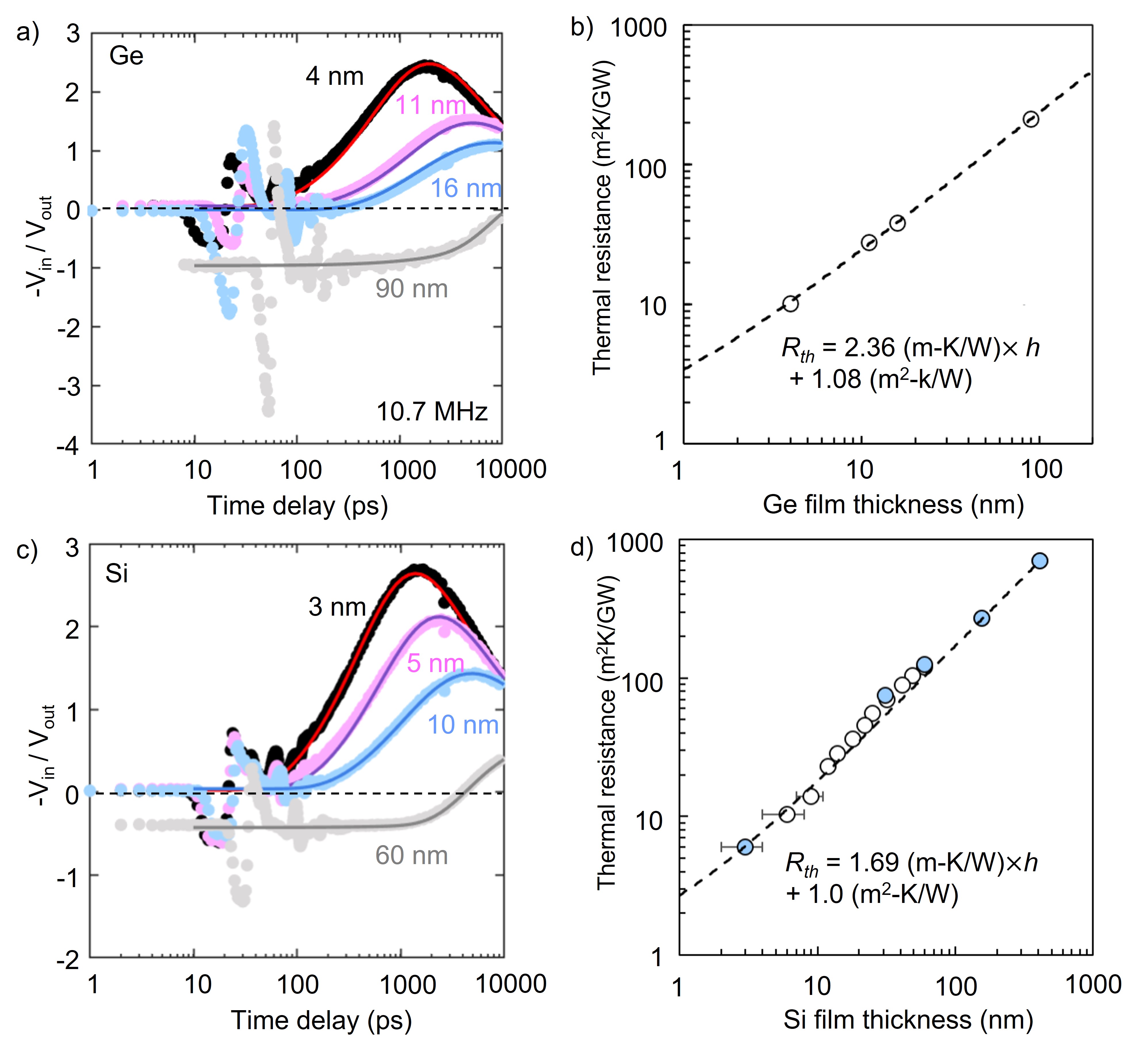}
    \caption{a) and c) The front/back TDTR data of amorphous Ge and amorphous Si as a function of film thickness. The circles are the experimental data. The lines are the best-fit predictions by the thermal model. b) The thermal resistance of amorphous Ge as a function of film thickness fit by front/front measurement. The observed linear relationship between thermal resistance and film thickness is consistent with a thickness independent thermal conductivity for a-Ge of 0.4 W/(m-K) and an Al/a-Ge interface conductance of $\sim$1850 MW/(m$^2$-K). d) The thermal resistance of amorphous Si layers as a function of film thickness. Open circles denote measurements on the a-Si wedge sample, while filled markers denote measurements of samples with a homogeneous a-Si film thickness. The observed linear relationship between thermal resistance and film thickness suggests a thickness-independent thermal conductivity for a-Si of ~$0.6$ W/(m-K) and an Al/a-Si interface conductance of $\sim$2000 MW/(m$^2$-K).  \textcolor{black}{A version of FIG 4 a) c) with a linear time-scale is shown in Supplementary Materials in Figure S4. } }
     \label{fb_resistance}
\end{figure*}

\noindent
\\
All measurements were done with a laser wavelength of 783 nm. 
The pump beam is modulated at a frequency of 10.7 MHz by an electro-optic modulator (Conoptics 350-160 with model 25D driver). The probe beam is modulated to 200 Hz by a mechanical chopper.
The pump beam goes through a mechanical delay stage (Akribis Systems DGL150), which allows a maximum of 13 ns delay relative to the probe pulses.
We measure the thermal response of the top Al layer at delay times between -30 ps to 9000 ps. 
A more detailed description of the experimental setup, optics, and technical discussions of our pump/probe system can be found in Ref \cite{gomez2020high}.

\noindent
\\
The data analysis of the front/front measurements was done with the well-established thermal model for TDTR \cite{cahill2004analysis,jiang2018tutorial}, whereas the front/back data were analyzed with a modified thermal model with different boundary conditions. The interface conductance between the Al and amorphous layers, and the thermal conductivity of the amorphous layers were treated as fit parameters. The heat capacity of amorphous Ge and amorphous Si are set to be 1.76 \cite{okhotinthermophysical} and 2.1 MJ/(m$^3$-K) \cite{zink2006thermal}. The thermal conductivity of Al is measured to be 120 W/(m-K).
\textcolor{black}{
The analytical solution used to analyze the front/back data is described in the Supplementary Materials and Ref. \cite{peng2021thermal}}.
%
%
%
%
\section{Thickness dependence of thermal conductivity of amorphous Si and Ge} \label{Results}
\noindent 
The thermal conductivities of amorphous Ge and Si films with varying thicknesses were measured with both the front/front and front/back TDTR configuration. 
The front/back results are summarized in FIG. \ref{fb_resistance}a and c. 
The in-phase and out-of-phase signals of the amorphous Ge as a function of modulation frequency for various film thicknesses in a front/back configuration are shown in Figure S3 in the Supplementary Materials. The circles are experimental data and the dashed lines are predictions by the thermal model. 
Both front/front and front/back measurements yield the same thermal conductivity and interface conductance. The best-fit thermal conductivity values for all samples are shown in Table S1 and S2 in the Supplementary Materials. %

\noindent
\\
We report the results of our experiments in FIG. \ref{fb_resistance}. Our measurements are sensitive to the total thermal resistance between Al layers, which includes both the thermal boundary resistance of the metal/amorphous-layer interface and the thermal resistance of the amorphous layer itself. The error bars in FIG. \ref{fb_resistance}b and \ref{fb_resistance}d represent $\sim$ 10\% uncertainty in thickness that arises from the roughness of the amorphous layer, speed of sound, etc. The uncertainty of the layer thickness has a negligible effect on our best fit value for the total thermal resistance between Al layers. However, the $\sim$ 10\% in film thickness results in ~$\sim$ 10\% uncertainty in the best-fit value for the thermal conductivity of each sample. 

\noindent
\\
The boundary resistance of the metal/amorphous-layer interface and the thermal resistance of the amorphous-layer will add in series.  Therefore, the total thermal resistance $R_{th}$ is 
\begin{equation}
    R_{th} = \frac{h}{\Lambda} + \frac{2}{G},
\end{equation}
where $h$ is the thickness of the amorphous layer, $\Lambda$ is the thermal conductivity of the amorphous layer, and $G$ is the thermal conductance of the Al/amorphous-layer interface.  If the thermal conductivity is independent of film thickness, Eq. (1) is a linear curve. Our measurements are in good agreement with Eq. (1) with a thickness independent ${\Lambda}$ = 0.4 and 0.6 W/(m-K) for a-Ge and a-Si, respectively. 

\noindent
\\
If long MFPs propagons could carry significant heat in our a-Si and a-Ge samples, we would expect the thermal conductivity to increase as a function of film thickness. This is because the thermal conductivity per propagation is proportional to the mean free path of the propagation. Since the mean-free path cannot be greater than the film thickness, the heat-carried by long-mean-path propagons should be suppressed as film thickness decreases. Our films do not have a thickness-dependent thermal conductivity. 

\noindent
\\
As illustrated in FIG. \ref{fb_resistance}b and FIG. \ref{fb_resistance}d, the thermal resistance of the a-Ge and a-Si show little evidence of size effect from 4 nm to 90 nm for amorphous Ge and 3nm to 400nm for amorphous Si. There may be a small change in the slope of the a-Si layer's thermal resistance vs. thickness near $\sim$30 nm.  But the deviation from a linear curve is small. Therefore, we conclude that propagons with mean-free-paths longer than a few nanometers carry a negligible amount of heat in a-Ge and a-Si prepared via RF sputtering at room temperature. 

%

\noindent
\\


%
%

\section{Summary and outlook}
\noindent
We report the results of thermal conductivity measurements of a-Si and a-Ge as a function of film thickness. The thermal conductivity was measured by traditional front/front TDTR experiments, as well a front/back TDTR method. The front/back TDTR geometry is a laser flash method with picosecond temporal resolution capable of accurately measuring the thermal conductivity of nanoscale layers. We observe a thickness independent thermal conductivity for a-Ge and a-Si of $0.4$ W/(m-K) and $0.6$ W/(m-K), respectively. The thickness independent thermal conductivity indicates that heat is primarily carried by diffusons and/or propagons with mean free paths shorter than a few nanometers. In addition to providing insight on the fundamental mechanisms responsible for heat transfer in amorphous materials, our study demonstrates that nanoscale laser flash experiments are an effective tool for characterizing thermal transport in nanoscale heterostructures. 
\section*{Supplementary Materials}
See the supplementary material for details on thermal model calculations, an analysis of how experimental signals depend on thermal properties and thickness of each layer, thicknesses and thermal parameters used calculation of TDTR signals, and frequency-dependent TDTR data.
\section*{Acknowledgement}
\noindent
This research was supported as part of ULTRA, an Energy Frontier Research Center funded by the U.S. Department of Energy (DOE), Office of Science, Basic Energy Sciences (BES), under Award \# DE-SC0021230 (thermal modelling).

\bibliography{apssamp}

\end{document}